\documentstyle[12pt]{article}

\begin{document}

\title{\bf Relativistic Causal Newton Gravity Law}

\author{Yury M. Zinoviev\thanks{This work was supported in part
by the Russian Foundation for Basic Research (Grant No. 12-01-00094), the
Program for Supporting Leading Scientific Schools (Grant No. 4612.2012.1)
and the RAS Program "Fundamental Problems of Nonlinear Mechanics."}}

\date{}
\maketitle

Steklov Mathematical Institute, Gubkin Street 8, 119991, Moscow,
Russia,

e - mail: zinoviev@mi.ras.ru

\vskip 0.5cm

\noindent {\bf Abstract.} The equations of the relativistic causal
Newton gravity law for the planets of the solar system are studied in
the approximation when the Sun rests at the coordinates origin and the
planets do not interact between each other.

\vskip 0.5cm

\section{Introduction}
\setcounter{equation}{0}

The Newton gravity law requires the instant propagation of the force
action. The special relativity requires that the propagation speed
does not exceed the speed of light. If the propagation speed is
independent of the gravitating body speed, then it is equal to that of
light. The special relativity requires also the gravity law covariance
under Lorentz transformations. Poincar\'e \cite{1} tried to find such a
modification of the Newton gravity law. (Poincar\'e considered two
mathematical problems in XX century as principal: "to create the
mathematical basis for the quantum physics and for the relativity
theory.") The gravity forces of two physical points should depend not on
its simultaneous positions and speeds but on the positions and the speeds
at the time moments which differ from each other in the time interval
needed for light covering the distance between the physical points. The
gravity force acting on one physical point may depend also on the
acceleration of another physical point at the delayed time moment.
The relativistic Newton gravity law was proposed in the paper
\cite{2}. This law for the two physical points has the form
\begin{equation}
\label{1.1} \frac{d}{dt} \left( \left( 1 - c^{- 2}\Bigl| \frac{d{\bf
x}_{k}}{dt}\Bigr|^{2}\right)^{- 1/2} \frac{dx_{k}^{\mu}}{dt} \right)
= - \eta^{\mu \mu} \sum_{\nu \, =\, 0}^{3}
c^{- 1}\frac{dx_{k}^{\nu}}{dt} F_{j;\mu
\nu}(x_{k},x_{j}),
\end{equation}
$j,k = 1,2$, $j \neq k$, $\mu = 0,...,3$. The world line
$x_{k}^{\mu}(t)$ satisfies the condition $x_{k}^{0}(t) = ct$; $c$ is
the speed of light; the diagonal $4\times 4$ - matrix
$\eta^{\mu \nu} = \eta_{\mu \nu}$, $\eta^{00} = -
\eta^{11} = - \eta^{22} = - \eta^{33} = 1$; the strength
$F_{j;\mu \nu}(x_{k},x_{j})$ is expressed through the vector potential
\begin{equation}
\label{1.2} F_{j;\mu \nu}(x_{k},x_{j}) = \frac{\partial
A_{j;\nu}(x_{k},x_{j})}{\partial x_{k}^{\mu}} - \frac{\partial
A_{j;\mu}(x_{k},x_{j})}{\partial x_{k}^{\nu}},
\end{equation}
\begin{equation}
\label{1.3} A_{j;\mu}(x_{k}, x_{j}) = \eta_{\mu \mu} m_{j}G\left(
\frac{d}{dt^{\prime}} x_{j}^{\mu}(t^{\prime})\right) \left( c|{\bf
x}_{k} - {\bf x}_{j}(t^{\prime})| - \sum_{i\, =\, 1}^{3} (x_{k}^{i}
- x_{j}^{i}(t^{\prime})) \frac{d}{dt^{\prime}}
x_{j}^{i}(t^{\prime})\right)^{- 1},
\end{equation}
$$
t^{\prime} = c^{- 1}(x_{k}^{0} - |{\bf x}_{k} - {\bf
x}_{j}(t^{\prime})|), \, \, j,k = 1,2, \, \, j \neq k;
$$
the gravitation constant $G = (6.673 \pm 0.003)\cdot 10^{-
11}m^{3}kg^{- 1}s^{- 2}$ and $m_{j}$ is the $j$ body mass. For a
resting body world line ($x_{j}^{0} (t) = ct$ and the vector ${\bf
x}_{j} (t)$ is constant) the vector potential (\ref{1.3}) coincides
with the Coulomb vector potential
\begin{equation}
\label{1.4} A_{j;0}(x_{k}, x_{j}) = m_{j}G|{\bf x}_{k} - {\bf
x}_{j}(c^{- 1}x_{k}^{0})|^{- 1},\, \, A_{j;i}(x_{k},x_{j}) = 0, \,
\, i = 1,2,3.
\end{equation}
If the velocities of bodies are small enough to neglect their
squares compared with the square of the light speed and it is
possible to neglect also the time interval $c^{- 1}|{\bf x}_{k} -
{\bf x}_{j}(t^{\prime})|$, then the vector potential (\ref{1.3}) is
nearly equal to the Coulomb vector potential (\ref{1.4}). The vector
potential (\ref{1.3}) was proposed by Li\'enard (1898) and Wiechert
(1900) as the generalization of the Coulomb vector potential
(\ref{1.4}). The substitution of the Coulomb vector potential
(\ref{1.4}) into the right-hand side of the equation (\ref{1.1}) for
$\mu = 1,2,3$ yields the right-hand side of the Newton gravity law
equations. The equation (\ref{1.1}) multiplied by
$\left( 1 - c^{- 2}|d{\bf x}_{k}/dt|^{2}\right)^{- 1/2}$
transforms as the vector. The equations (\ref{1.1}), (\ref{1.2}) with the
Li\'enard - Wiechert vector potential (\ref{1.3}) are the relativistic
version of the Newton gravity law equations.

Sommerfeld (\cite{6}, Sec. 38): "The question may arise: what is the
relativistic form of the Newton gravity law? If the law is supposed
to have a vector form, this question is wrong. The gravitational
field is not a vector field. It has the incomparably
complicated tensor structure." The Newton gravity law equations and
the equations (\ref{1.1}) - (\ref{1.3}) define the interactions. The
body interacts only with another body. If two bodies create the
common gravitational field with the vector potential $A_{1;\mu}(x,
x_{1}) + A_{2;\mu}(x, x_{2})$, any body should interact with itself
and we obtain the infinity in the equations (\ref{1.3}),
(\ref{1.4}) at $x_{k} = x_{j}$. The notion of gravitational field with
the vector potential $A_{1;\mu}(x, x_{1}) + A_{2;\mu}(x, x_{2})$ is not
compatible with the Newton gravity law and with the relativistic Newton
gravity law (\ref{1.1}) - (\ref{1.3}).

The delay $c^{- 1}|{\bf x}_{k} - {\bf x}_{j}(t^{\prime})|$ in the
relation (\ref{1.3}) provides the causality condition according to
which some event in the system can influence the evolution of the
system in the future only and can not influence the behavior of the
system in the past, in the time preceding the given event. The delay
$c^{- 1}|{\bf x}_{k} - {\bf x}_{j}(t^{\prime})|$ in the relation
(\ref{1.3}) is very important: one celestial body is a good distance
off another celestial body. Poincar\'e \cite{1}: "It turned out to be
necessary to consider this hypothesis more attentively and to study the
changes it makes in the gravity laws in particular. First, it obviously
enables us to suppose that the gravity forces propagate not instantly
but at the speed of light." The general relativity does not take into
account the causality condition and the delay.

$99.87\%$ of the total mass of the solar system belongs to the Sun.
We consider the relativistic causal Newton gravity law equations
\cite{2} for the planets of the solar system in the natural
approximation when the Sun rests at the coordinates origin and the
planets do not interact between each other. In this approximation
the problem of planet relativistic motion was solved in the paper
\cite{2}. The planet orbits were given by the formulas which differ
from the formulas defining the ellipses in the precession coefficients
only. The precession coefficients for the solar system planets are
practically equal to one. The similar orbits with another precession
coefficients are considered in the general relativity (\cite{4}, Chap.
40, Sec. 40.5, relations (40.17), (40.18)). In the beginning of the
XVII century Johannes Kepler by making use of Tycho Brahe (1546 - 1601)
astronomical observations found that the planet orbits are elliptic in
the coordinate system where the Sun rests (Nicolaus Copernicus (1543)).
The intensive astronomic observations from the middle of the XIX century
and the radio-location after 1966 discovered the advances of orbit
perihelion for different planets.

In the general relativity the observed value for the Mercury's
perihelion advance is obtained by means of addition the advance
of Mercury's perihelion (\cite{4}, Chap. 40, Sec. 40.5, Appendix 40.3)
calculated in the Newton gravity theory and the advance of Mercury's
perihelion calculated for the orbits (\cite{4}, Chap. 40, Sec. 40.5,
relations (40.17), (40.18)). The orbits (\cite{4}, Chap. 40, Sec. 40.5,
relations (40.17), (40.18)) are the approximate solutions of the geodesic
equation for the chosen metrics (\cite{4}, Chap. 40, Sec. 40.1, relation
(40.3)). It is not obvious that we can add the advance of Mercury's
perihelion obtained for the orbits (\cite{4}, Chap. 40, Sec. 40.5,
Appendix 40.3) and for the orbits (\cite{4}, Chap. 40, Sec. 40.5,
relations (40.17), (40.18)). It seems natural to obtain the advance of
Mercury's perihelion, observed from the Earth, by making use of the
Mercury and Earth orbits (\cite{4}, Chap. 40, Sec. 40.5, relations (40.17),
(40.18)) calculated without Newton gravity theory. In order to calculate
the advance of Mercury's perihelion we need to know also the time
dependence of the orbit (\cite{4}, Chap. 40, Sec. 40.5, relations (40.17),
(40.18)) radius. In this paper we study the similar orbits of the
relativistic causal Newton gravity law \cite{2}. We shall show in this
paper that the value of the Mercury's perihelion advance, observed from
the Earth, depends on the perihelion angles of the Mercury and Earth
orbits. The perihelion angle of the planet's orbit depends on the
planet's perihelion point due to the precession coefficient in the
planet's orbit formula. For the experimental verification of the
equations (\ref{1.1}) - (\ref{1.3}) the perihelion angles of the Mercury
and Earth orbits are needed.

\section{Causal Coulomb and Newton laws}
\setcounter{equation}{0}

The relativistic Lagrange law is the particular case of
the relativistic Newton second law
\begin{equation}
\label{1.17} mc\frac{dt}{ds} \frac{d}{dt} \left( \frac{dt}{ds}
\frac{dx^{\mu}}{dt} \right) + qc^{- 1} \sum_{k\, =\, 0}^{N}
\, \sum_{\alpha_{1}, ..., \alpha_{k} \, =\, 0}^{3} \eta^{\mu \mu}
F_{\mu \alpha_{1} \cdots \alpha_{k}}(x)\frac{dt}{ds}
\frac{dx^{\alpha_{1}}}{dt} \cdots \frac{dt}{ds}
\frac{dx^{\alpha_{k}}}{dt} = 0,
\end{equation}
$$
\frac{dt}{ds} = \left( c^{2} - |{\bf v}|^{2}\right)^{- 1/2},
\, \, v^{i} = \frac{dx^{i}}{dt}, \, \, i = 1,2,3.
$$
where $\mu = 0,...,3$ and the world line $x^{\mu}(t)$
satisfies the condition: $x^{0}(t)= ct$. The force is the polynomial
of the speed in the equation (\ref{1.17}). It is necessary to define
the series convergence for the force as an infinite series of the speed.
The second relation (\ref{1.17}) implies the identities
\begin{equation}
\label{1.19} \sum_{\alpha \, =\, 0}^{3} \eta_{\alpha \alpha} \left(
\frac{dt}{ds} \frac{dx^{\alpha}}{dt} \right)^{2} = 1, \, \,
\sum_{\alpha \, =\, 0}^{3} \eta_{\alpha \alpha} \frac{dt}{ds}
\frac{dx^{\alpha}}{dt} \frac{dt}{ds} \frac{d}{dt} \left(
\frac{dt}{ds} \frac{dx^{\alpha}}{dt} \right) = 0.
\end{equation}
The equation (\ref{1.17}) and the second identity (\ref{1.19}) imply
\begin{equation}
\label{1.20} \sum_{k\, =\, 0}^{N} \sum_{\alpha_{1}, ..., \alpha_{k +
1} \, =\, 0}^{3} F_{\alpha_{1} \cdots \alpha_{k + 1}}(x)
\frac{dt}{ds} \frac{dx^{\alpha_{1}}}{dt} \cdots \frac{dt}{ds}
\frac{dx^{\alpha_{k + 1}}}{dt} = 0.
\end{equation}
Let the functions $F_{\alpha_{1} \cdots \alpha_{k + 1}}(x)$ satisfy
the equation (\ref{1.20}). Then three equations (\ref{1.17}) for
$\mu = 1,2,3$ are independent
\begin{eqnarray}
\label{1.21} m\frac{d}{dt} \left( (1 - c^{- 2}|{\bf v}|^{2})^{-
1/2}v^{i}\right) - qc^{- 1}\sum_{k\, =\, 0}^{N} (c^{2} - |{\bf
v}|^{2})^{- (k - 1)/2} \nonumber \\ \times \left( \sum_{\alpha_{1},
..., \alpha_{k} \, =\, 0}^{3} F_{i\alpha_{1} \cdots
\alpha_{k}}(x)\frac{dx^{\alpha_{1}}}{dt} \cdots
\frac{dx^{\alpha_{k}}}{dt} \right) = 0, \, \, i = 1,2,3.
\end{eqnarray}
The following lemma is proved in the paper \cite{2}.

\noindent {\bf Lemma}. {\it Let there exist a Lagrange function}
$L({\bf x},{\bf v},t)$ {\it such that for any world line}
$x^{\mu}(t)$, $x^{0}(t) = ct$, {\it the relation}
\begin{eqnarray}
\label{1.22} \frac{d}{dt} \frac{\partial L}{\partial v^{i}} -
\frac{\partial L}{\partial x^{i}} = m\frac{d}{dt} \left( (1 - c^{-
2}|{\bf v}|^{2})^{- 1/2}v^{i}\right) - qc^{- 1}\sum_{k\, =\, 0}^{N}
\nonumber \\ (c^{2} - |{\bf v}|^{2})^{- (k - 1)/2}
\sum_{\alpha_{1}, ..., \alpha_{k} \, =\, 0}^{3}F_{i\alpha_{1} \cdots
\alpha_{k}}(x)\frac{dx^{\alpha_{1}}}{dt} \cdots
\frac{dx^{\alpha_{k}}}{dt}
\end{eqnarray}
{\it holds for any} $i = 1,2,3$. {\it Then the Lagrange function has
the form}
\begin{equation}
\label{1.23} L({\bf x},{\bf v},t) = - mc^{2}(1 - c^{- 2}|{\bf
v}|^{2})^{1/2} + q \sum_{i\, =\, 1}^{3} A_{i}({\bf
x},t)c^{- 1}v^{i} + qA_{0}({\bf x},t)
\end{equation}
{\it and the coefficients} $F_{i\alpha_{1} \cdots
\alpha_{k}}(x)$ {\it in the equations} (\ref{1.21}) {\it are}
\begin{equation}
\label{1.24} F_{i\alpha_{1} \cdots \alpha_{k}}(x) = 0,\, \, k \neq
1,\, i = 1,2,3,\, \alpha_{1}, ...,\alpha_{k} = 0,...,3,
\end{equation}
\begin{eqnarray}
\label{1.25} F_{ij}(x) = \frac{\partial A_{j}({\bf x},t)}{\partial
x^{i}} - \frac{\partial A_{i}({\bf x},t)}{\partial x^{j}},\, \, i,j
= 1,2,3, \nonumber
\\ F_{i0}(x) = \frac{\partial A_{0}({\bf x},t)}{\partial
x^{i}} - \frac{1}{c} \frac{\partial A_{i}({\bf x},t)}{\partial t},\,
\, i = 1,2,3.
\end{eqnarray}
We define the coefficients
\begin{equation}
\label{1.26} F_{00} (x) = 0,\, \, F_{0i} (x) = - F_{i0} (x),\, \, i = 1,2,3.
\end{equation}
Then the identity
\begin{equation}
\label{1.27} \sum_{\alpha, \beta \, =\, 0}^{3} F_{\alpha
\beta}(x)\frac{dt}{ds} \frac{dx^{\alpha}}{dt} \frac{dt}{ds}
\frac{dx^{\beta}}{dt} = 0
\end{equation}
of the type (\ref{1.20}) holds. By making use of the
second identity (\ref{1.19}) and the relations (\ref{1.25}) -
(\ref{1.27}) we can rewrite the equation (\ref{1.21}) with the
coefficients (\ref{1.24}), (\ref{1.25}) as the relativistic Newton
second law with Lorentz force
\begin{eqnarray}
\label{1.28} mc\frac{dt}{ds} \frac{d}{dt} \left( \frac{dt}{ds}
\frac{dx^{\mu}}{dt} \right) = - q \eta^{\mu \mu} \sum_{\nu
\, =\, 0}^{3} F_{\mu \nu}(x)c^{- 1} \frac{dt}{ds} \frac{dx^{\nu}}{dt},
\nonumber \\ F_{\mu \nu}(x) = \frac{\partial A_{\nu}({\bf
x},t)}{\partial x^{\mu}} - \frac{\partial A_{\mu}({\bf
x},t)}{\partial x^{\nu}}, \, \, \mu, \nu = 0,...,3.
\end{eqnarray}
For the relativistic Lagrange law the interaction is defined by the
product of the charge $q$ and the external vector potential
$A_{\mu}({\bf x},t)$.

Let a distribution $e_{0} (x) \in S^{\prime} ({\bf R}^{4})$ with
support in the closed upper light cone be a fundamental solution of
the wave equation
\begin{equation}
\label{112.33} - (\partial_{x},\partial_{x})
e_{0} (x) = \delta (x), \, \, (\partial_{x}, \partial_{x}) =
\left( \frac{\partial}{\partial x^{0}} \right)^{2} - \sum_{i\, =\, 1}^{3}
\left( \frac{\partial}{\partial x^{i}} \right)^{2}.
\end{equation}
We prove the uniqueness of the equation (\ref{112.33}) solution in
the class of distributions with supports in the closed upper
light cone. Let the equation (\ref{112.33}) have two solutions
$e_{0}^{(1)} (x)$, $e_{0}^{(2)} (x)$. Since its
supports lie in the closed upper light cone, the convolution is
defined. Now the convolution commutativity
\begin{equation}
\label{112.332} \int d^{4}x d^{4}y e_{0}^{(2)} (x - y)
e_{0}^{(1)} (y) \phi (x) = \int d^{4}x d^{4}y
e_{0}^{(1)} (x) e_{0}^{(2)} (y) \phi (x + y)
\end{equation}
implies these distributions coincidence:
\begin{equation}
\label{112.331} e_{0}^{(j)} (x) = - (\partial_{x},
\partial_{x}) \int d^{4}y e_{0}^{(k)} (x - y)
e_{0}^{(j)} (y),
\end{equation}
$j,k = 1,2$, $j \neq k$. Due to the book (\cite{3}, Sect. 30)
this unique causal distribution is
\begin{equation}
\label{112.34} e_{0}(x) = -\, (2\pi)^{- 1} \theta (x^{0})\delta
((x,x)),
\end{equation}
$$
(x,y) = x^{0}y^{0} - \sum_{k\, =\, 1}^{3} x^{k}y^{k}, \, \,
\theta (x) = \left\{ {1, \hskip 0,5cm x \geq 0,} \atop
{0, \hskip 0,5cm x < 0.} \right.
$$

The relativistic causal Coulomb law is given by the equations of the type
(\ref{1.28})
\begin{eqnarray}
\label{12.1} m_{k} \frac{d}{dt} \left( \left( 1 - c^{- 2}\Bigl|
\frac{d{\bf x}_{k}}{dt}\Bigr|^{2} \right)^{- 1/2}
\frac{dx_{k}^{\mu}}{dt} \right) = - q_{k} \eta^{\mu \mu}
\sum_{\nu \, =\, 0}^{3} c^{- 1}\frac{dx_{k}^{\nu}}{dt}
F_{j;\mu \nu}(x_{k},x_{j}),
\end{eqnarray}
$j,k = 1,2$, $j \neq k$,  where the strength
$F_{j;\mu \nu}(x_{k},x_{j})$ is given by the relation (\ref{1.2})
with the Li\'enard - Wiechert vector potential of the type (\ref{1.3})
$$
A_{j; \mu}(x_{k},x_{j}) = - \, 4\pi q_{j}K \sum_{\nu \,
= \, 0}^{3} \eta_{\mu \nu} \int dt e_{0}(x_{k} - x_{j}(t))
\frac{dx_{j}^{\nu }(t)}{dt} =
$$
\begin{equation}
\label{12.4}
- \, q_{j}K \eta_{\mu \mu} \left( \frac{d}{dt}
x_{j}^{\mu}(t)\right) \left( c|{\bf x}_{k} - {\bf x}_{j}(t)| -
\sum_{i\, =\, 1}^{3} (x_{k}^{i} - x_{j}^{i}(t)) \frac{d}{dt}
x_{j}^{i}(t)\right)^{- 1} \Biggl|_{t = t(0)},
\end{equation}
$$
x_{k}^{0} - ct(0) = |{\bf x}_{k} - {\bf x}_{j}(t(0))|.
$$
Here $K$ is the constant of the causal electromagnetic interaction
for two particles with the charges $q_{j}$. The support of the
distribution (\ref{112.34}) lies in the upper light cone boundary.
The interaction speed is equal to that of light. It is easy to prove
the second relation (\ref{12.4}) by making change of the integration
variable
\begin{equation}
\label{12.17} x_{k}^{0} - ct(r) = (|{\bf x}_{k} - {\bf
x}_{j}(t(r))|^{2} + r)^{1/2}.
\end{equation}
For $r = 0$ the relation (\ref{12.17}) coincides with the third relation
(\ref{12.4}).

The equations (\ref{12.1}), (\ref{1.2}), (\ref{12.4}) are the relativistic
causal version of the Coulomb law. The Lorentz invariant distribution
(\ref{112.34}) defines the delay. The Lorentz invariant solutions of the
equation (\ref{112.33}) are described in the paper \cite{2}. By making use
of these solutions it is possible to describe the Lorentz covariant
equations of the type (\ref{12.1}), (\ref{1.2}), (\ref{12.4}). The
equations (\ref{12.1}), (\ref{1.2}), (\ref{12.4}) are Lorentz covariant
and causal due to the distribution (\ref{112.34}). The quantum version
of the equations (\ref{12.1}), (\ref{1.2}), (\ref{12.4}) is defined in
the paper \cite{8}. The solutions of these causal equations do not
contain the diverging integrals similar to the diverging integrals of
the quantum electrodynamics.

For a world line $x_{j}^{\mu}(t)$ we define the vector proportional to
$-  \eta^{\mu \mu} (\partial_{x},\partial_{x})A_{j; \mu}(x,x_{j})$
$$
J^{\mu}(x,x_{j}) =  - (\partial_{x},\partial_{x})
\int dt e_{0}(x - x_{j}(t)) \frac{dx_{j}^{\mu }(t)}{dt} =
\int dt \delta (x - x_{j}(t)) \frac{dx_{j}^{\mu }(t)}{dt} =
$$
\begin{equation}
\label{12.10}  \left( \frac{d}{dx^{0}}
x_{j}^{\mu} \left( c^{- 1}x^{0} \right) \right) \delta \left( {\bf
x} - {\bf x}_{j} \left( c^{- 1} x^{0} \right) \right), \mu = 0,...,3.
\end{equation}
The condition $x_{j}^{0}(t) = ct$ implies the continuity equation
\begin{eqnarray}
\label{12.121} \frac{\partial}{\partial x^{0}} J^{0}(x,x_{j}) = -
\sum_{i\, =\, 1}^{3} \left( \frac{d}{dx^{0}} x_{j}^{i} \left( c^{-
1}x^{0} \right) \right) \frac{\partial}{\partial x^{i}} \delta
\left( {\bf x} - {\bf
x}_{j} \left( c^{- 1} x_{k}^{0} \right) \right), \nonumber \\
\frac{\partial}{\partial x^{i}} J^{i}(x,x_{j}) = \left(
\frac{d}{dx^{0}} x_{j}^{i} \left( c^{- 1}x^{0} \right) \right)
\frac{\partial}{\partial x^{i}} \delta \left( {\bf x} - {\bf x}_{j}
\left( c^{- 1} x^{0} \right) \right),\, i = 1,2,3,
\end{eqnarray}
\begin{equation}
\label{12.12} \sum_{\mu \, =\, 0}^{3} \frac{\partial}{\partial
x^{\mu}} J^{\mu}(x,x_{j}) = 0.
\end{equation}
The integration of the relation
\begin{equation}
\label{12.13} e_{0}(x - x_{j}(t)) \frac{dx_{j}^{\mu }(t)}{dt} = \int
d^{4}y e_{0}(x - y) \delta (y - x_{j}(t))
\frac{dx_{j}^{\mu}(t)}{dt}
\end{equation}
along the world line $x_{j}^{\mu}(t)$ yields
\begin{equation}
\label{12.14} \int dt e_{0}(x - x_{j}(t)) \frac{dx_{j}^{\mu
}(t)}{dt} = \int d^{4}y e_{0}(x - y) J^{\mu}(y,x_{j}).
\end{equation}
The relations (\ref{12.12}), (\ref{12.14}) imply the gauge condition
for the vector potential (\ref{12.4})
\begin{equation}
\label{12.16} \sum_{\mu \, =\, 0}^{3} \eta^{\mu \mu}
\frac{\partial}{\partial x^{\mu}} A_{j;\mu}(x,x_{j}) = 0.
\end{equation}
Due to the gauge condition (\ref{12.16}) the tensor (\ref{1.2}),
(\ref{12.4}) satisfies Maxwell equations with the current
proportional to the current (\ref{12.10}).

The substitution $K = - G$ and two positive or two negative
gravitational masses $q_{1} = \pm m_{1}$, $q_{2} = \pm m_{2}$
into the equations (\ref{12.1}), (\ref{1.2}), (\ref{12.4})
yields the relativistic causal Newton gravity law (\ref{1.1}) -
(\ref{1.3}). By changing the constants $K = - G$, $q_{1} = \pm m_{1}$,
$q_{2} = \pm m_{2}$ in the equations from the paper \cite{8} we have
the quantum version of the equations (\ref{1.1}) - (\ref{1.3}). The
substitution $K = - G$ and also one positive and one negative
gravitational masses $q_{1} = \pm m_{1}$, $q_{2} = \mp m_{2}$ into
the equations (\ref{12.1}), (\ref{1.2}), (\ref{12.4}) yields the
galaxies scattering with an acceleration. Einstein \cite{5}:
"The theoretical physicists studying the problems of the general
relativity can hardly doubt now that the gravitational and
electromagnetic fields should have the same nature."

\section{Advance of Mercury's perihelion}
\setcounter{equation}{0}

Due to the paper \cite{2} the relativistic causal Newton gravity law for the
solar system has the form
\begin{equation}
\label{2.1} \frac{d}{dt} \left( \left( 1 - c^{- 2}\Bigl| \frac{d{\bf
x}_{k}}{dt}\Bigr|^{2} \right)^{- 1/2} \frac{dx_{k}^{\mu}}{dt}
\right) = - \eta^{\mu \mu} \sum_{\nu \, =\, 0}^{3}
c^{- 1}\frac{dx_{k}^{\nu}}{dt} \sum_{j\, =\, 1,...,10,\, j \neq k} F_{j;\mu
\nu}(x_{k},x_{j}).
\end{equation}
We give the number $k = 1$ for Mercury, the number
$k = 2$ for Venus, the number $k = 3$ for the Earth, the number $k = 4$ for
Mars, the number $k = 5$ for Jupiter, the number $k = 6$ for Saturn, the
number $k = 7$ for Uranus, the number $k = 8$ for Neptune, the number
$k = 9$ for Pluto and the number $k = 10$ for the Sun.

$99.87\%$ of the total mass of the solar system belongs to the Sun.
We consider the Sun resting at the coordinates origin (Nicolaus Copernicus
(1543)). Substituting the Sun world line $x_{10}^{0}(t) = ct$,
$x_{10}^{i}(t) = 0$, $i = 1,2,3$, into the equalities (\ref{1.2}),
(\ref{1.3}) we have
\begin{equation}
\label{2.8} F_{10;ij}(x;x_{10}) = 0,\, \, i,j = 1,2,3,\, \,
F_{10;i0}(x;x_{10}) = - m_{10}G|{\bf x}|^{- 3}x^{i},\, \, i = 1,2,3.
\end{equation}
Substituting the Sun world line $x_{10}^{0}(t) = ct$, $x_{10}^{i}(t) =
0$, $i = 1,2,3$, and the elliptic orbits into the expressions
(\ref{1.2}) and (\ref{1.3}) it is possible to show that the values
of strengths $F_{10;i0}(x_{k};x_{10})$ considerably exceed the
values of strengths $F_{j;i \nu }(x_{k};x_{j})$ for any $k,j =
1,...,9$, $k \neq j$. It is possible to show also that the values of
strengths $F_{j;i \nu }(x_{10};x_{j})$ are negligible. We neglect
the action of any planet on all of the other planets and the Sun.
Then the Sun rests at the coordinates origin. Due to the relations
(\ref{2.8}) in the coordinates system where the Sun rests at the
coordinates origin the first nine equations (\ref{2.1}) have the form
\begin{equation}
\label{2.9} \frac{d}{dt} \left( \left( 1 - c^{- 2}\Bigl| \frac{d{\bf
x}_{k}}{dt}\Bigr|^{2}\right)^{- 1/2} \frac{dx_{k}^{i}}{dt} \right) =
- m_{10}G|{\bf x}_{k}|^{- 3}x_{k}^{i},\, \, i = 1,2,3,\, \, k =
1,...,9
\end{equation}
It is shown in the paper \cite{2} that the following values
\begin{equation}
\label{2.10} M_{l}({\bf x}_{k}) = \sum_{i,j = 1}^{3} \epsilon_{ijl}
\left( x_{k}^{i}\frac{dx_{k}^{j}}{dt} -
x_{k}^{j}\frac{dx_{k}^{i}}{dt} \right) \left( 1 - \frac{1}{c^{2}}
\Bigl| \frac{d{\bf x}_{k}}{dt} \Bigr|^{2} \right)^{- 1/2},\, \, l =
1,2,3,
\end{equation}
\begin{equation}
\label{2.11} E({\bf x}_{k}) = c^{2}\left( 1 - \frac{1}{c^{2}} \Bigl|
\frac{d{\bf x}_{k}}{dt}\Bigr|^{2} \right)^{- 1/2} - m_{10}G|{\bf
x}_{k}|^{- 1},\, \, k = 1,...,9.
\end{equation}
are conserved for the equations (\ref{2.9}). The antisymmetric in
all indices tensor $\epsilon_{ijl}$ has the normalization
$\epsilon_{123} = 1$. The conservation of the vector (\ref{2.10}) is
the relativistic second Kepler law. The vector ${\bf x}_{k}$ is
orthogonal to the constant vector (\ref{2.10}). We introduce the
polar coordinates in the plane orthogonal to the vector (\ref{2.10})
\begin{equation}
\label{2.12} x_{\bot; k}^{1} (t) = r_{k}(t)\cos \phi_{k} (t),\, \,
x_{\bot; k}^{2} (t) = r_{k}(t)\sin \phi_{k} (t), \, \, k = 1,...,9.
\end{equation}

Let the constants (\ref{2.10}), (\ref{2.11}) satisfy the
inequalities
\begin{equation}
\label{2.13} c^{2}|{\bf M}({\bf x}_{k})|^{2} - m_{10}^{2}G^{2} > 0,
\end{equation}
\begin{equation}
\label{2.14} |{\bf M}({\bf x}_{k})|^{2}((E({\bf x}_{k}))^{2} -
c^{4}) + m_{10}^{2}G^{2}c^{2} > 0, \, \, k = 1,...,9.
\end{equation}
Due to the paper \cite{2} the equations (\ref{2.9}) have the
solutions
\begin{equation}
\label{2.15} \frac{p_{k}}{r_{k}(t)} = 1 + e_{k} \cos \left(
\gamma_{k} (\phi_{k} (t) - \phi_{k;0} )\right)
\end{equation}
where $\phi_{k;0} $ is the constant perihelion angle and the
constants
\begin{eqnarray}
\label{2.16} p_{k} = (c^{2}|{\bf M}({\bf x}_{k})|^{2} -
m_{10}^{2}G^{2})(m_{10}GE({\bf x}_{k}))^{- 1}, \nonumber \\
e_{k} = (c^{2}|{\bf M}({\bf x}_{k})|^{2}((E({\bf
x}_{k}))^{2} - c^{4}) + m_{10}^{2}G^{2}c^{4})^{1/2} (m_{10}GE({\bf
x}_{k}))^{- 1}, \nonumber \\
\gamma_{k} = (c^{2}|{\bf M}({\bf x}_{k})|^{2} -
m_{10}^{2}G^{2})^{1/2} (c|{\bf M}({\bf x}_{k})|)^{- 1}, \, \, k =
1,...,9.
\end{eqnarray}
The equations (\ref{2.15}) are the relativistic first Kepler law.
The orbit (\ref{2.15}) is not periodic in general. The substitution
of the vector (\ref{2.12}) with $r_{k}(t) = a_{k}$, $\phi_{k} (t) =
\omega_{k} (t - t_{k;0})$ in the equations (\ref{2.9}) yields the
relativistic third Kepler law:
\begin{equation}
\label{2.17} (1 - c^{- 2}a_{k}^{2}\omega_{k}^{2})^{-
1/2}a_{k}^{3}\omega_{k}^{2} = m_{10}G.
\end{equation}
The equations (\ref{2.15}) define the trajectory of motion but do
not define the time dependence of this trajectory. Let the constants
(\ref{2.10}), (\ref{2.11}) satisfy the inequality (\ref{2.14}) and
the inequalities
\begin{equation}
\label{2.18} (E({\bf x}_{k}))^{2} < c^{4}, \, \, k = 1,...,9.
\end{equation}
Due to the paper \cite{2} the equations (\ref{2.9}) have the
solutions with the constant parameter $\xi_{k;0} $
\begin{eqnarray}
\label{2.19} r_{k}(\xi_{k} ) = m_{10}GE({\bf x}_{k})(c^{4} - (E({\bf
x}_{k}))^{2})^{- 1} \nonumber \\ \times \Bigl( 1 + e_{k}\sin (
(c^{4} - (E({\bf x}_{k}))^{2})^{1/2}c^{- 1} \xi_{k} ) \Bigr),
\nonumber \\
t_{k}(\xi_{k}) = m_{10}G(E({\bf x}_{k}))^{2} c^{- 1}(c^{4} - (E({\bf
x}_{k}))^{2})^{- 3/2} \nonumber \\ \times \Bigl( c^{3}(E({\bf
x}_{k}))^{- 2} (c^{4} - (E({\bf x}_{k}))^{2})^{1/2}(\xi_{k} -
\xi_{k;0}) \nonumber \\ - e_{k}\cos ( (c^{4} - (E({\bf
x}_{k}))^{2})^{1/2}c^{- 1}\xi_{k} ) \Bigr), \, \, k = 1,...,9.
\end{eqnarray}

Let us express the constants in the equations (\ref{2.15}),
(\ref{2.19}) trough the astronomical orbit data. The orbit
eccentricities: $e_{1} = 0.21$, $e_{2} = 0.007$, $e_{3} = 0.017$,
$e_{4} = 0.093$, $e_{5} = 0.048$, $e_{6} = 0.056$, $e_{7} = 0.047$,
$e_{8} = 0.009$, $e_{9} = 0.249$. Therefore $0 < e_{k} < 1$, $k =
1,...,9$. Let us suppose $E({\bf x}_{k}) > 0$, $k = 1,...,9$. The
curve (\ref{2.15}) is an ellipse with a precession. The focus of
this ellipse is the coordinates origin. The major and minor "semi -
axes" are equal to
\begin{equation}
\label{2.20} a_{k} = p_{k}(1 - e_{k}^{2})^{- 1} = m_{10}GE({\bf
x}_{k})(c^{4} - (E({\bf x}_{k}))^{2})^{- 1},
\end{equation}
\begin{equation}
\label{2.21} b_{k} = a_{k}(1 - e_{k}^{2})^{1/2} = (c^{2}|{\bf
M}({\bf x}_{k})|^{2} - m_{10}^{2}G^{2})^{1/2} (c^{4} - (E({\bf
x}_{k}))^{2})^{- 1/2}, \, \, k = 1,...,9.
\end{equation}
The inequalities (\ref{2.13}) and $e_{k}^{2} < 1$ imply the
inequality (\ref{2.18}). Hence the Eqs. (\ref{2.19}) hold. For the
parameters $\xi_{k;\pm} = \pm (\pi /2)c(c^{4} - (E({\bf
x}_{k}))^{2})^{- 1/2}$ we have the extremal radii
\begin{equation}
\label{2.22} r_{k}(\xi_{k;\pm} ) =  m_{10}GE({\bf x}_{k})(c^{4} -
(E({\bf x}_{k}))^{2})^{- 1}(1 \pm e_{k}),\, \, k = 1,...,9.
\end{equation}
Hence, the "period" of the motion along the ellipse (\ref{2.15}) is
equal to
\begin{equation}
\label{2.24} T_{k} = 2|t_{k}(\xi_{k;+} ) - t_{k}(\xi_{k;-} )| = 2\pi
m_{10}Gc^{3}(c^{4} - (E({\bf x}_{k}))^{2})^{- 3/2}, \, \, k =
1,...,9.
\end{equation}
Let us define the mean "angular frequency" $\omega_{k} = 2\pi
T_{k}^{- 1}$. The relation (\ref{2.24}) implies
\begin{eqnarray}
\label{2.26} \omega_{k} = (c^{4} - (E({\bf
x}_{k}))^{2})^{3/2}(m_{10}Gc^{3})^{- 1},
\nonumber \\
(E({\bf x}_{k}))^{2} = c^{2}(c^{2} - (\omega_{k} m_{10}G)^{2/3}),\,
\, k = 1,...,9.
\end{eqnarray}
The substitution of the expression (\ref{2.26}) into the equality
(\ref{2.20}) yields
\begin{equation}
\label{2.27} m_{10}G = \omega_{k}^{2} a_{k}^{3} \left( 2^{- 1}(1 +
\sigma_{k} (1 - (2a_{k}\omega_{k} c^{- 1})^{2})^{1/2})\right)^{-
3/2}, \, \, \sigma_{k} = \pm 1, \, \, k = 1,...,9.
\end{equation}
Let $c \rightarrow \infty $. Then $m_{10}G = \omega_{k}^{2}
a_{k}^{3}(2^{- 1}(1 + \sigma_{k} ))^{- 3/2}$. For $\sigma_{k} = 1$
this expression agrees with the third Kepler law $m_{10}G =
\omega_{k}^{2} a_{k}^{3}$. Choosing $\sigma_{k} = 1$ in the relation
(\ref{2.27}) we get the "relativistic third Kepler law" for the
orbit (\ref{2.15})
\begin{equation}
\label{2.28}  \omega_{k}^{- 2} a_{k}^{- 3}m_{10}G = \left( 2^{- 1}(1 +
(1 - 4\omega_{k}^{2} a_{k}^{2} c^{- 2})^{1/2})\right)^{- 3/2}
\approx 1 + \frac{3}{2}  \omega_{k}^{2} a_{k}^{2}c^{- 2},\, \, k =
1,...,9.
\end{equation}
According to the book (\cite{4}, Chap. 25, Sec. 25.1, Appendix 25.1)
the values $\omega_{k}^{2} a_{k}^{3}c^{- 2} = 1477m$
for $k = 1,2,3,4,6$ (Mercury, Venus, the Earth, Mars and Saturn),
the values $\omega_{l}^{2} a_{l}^{3}c^{- 2} = 1478m$ for $l = 5,8$
(Jupiter and Neptune), the value $\omega_{7}^{2} a_{7}^{3}c^{- 2} = 1476m$
for Uranus, the value $\omega_{9}^{2} a_{9}^{3}c^{- 2} = 1469m$ for Pluto;
the major semi-axes   $a_{1} = 0.5791\cdot 10^{11}m$, $a_{2} = 1.0821\cdot
10^{11}m$, $a_{3} = 1.4960\cdot 10^{11}m$, $a_{4} = 2.2794\cdot
10^{11}m$, $a_{5} = 7.783\cdot 10^{11}m$, $a_{6} = 14.27\cdot
10^{11}m$, $a_{7} = 28.69\cdot 10^{11}m$, $a_{8} = 44.98\cdot
10^{11}m$, $a_{9} = 59.00\cdot 10^{11}m$. The values
$\omega_{k}^{2}a_{k}^{2} c^{- 2}= a_{k}^{- 1}\cdot \omega_{k}^{2}a_{k}^{3} c^{- 2}$,
$k = 1,...,9$, are negligible and therefore the Sun mass values (\ref{2.28})
obtained in the relativistic Kepler problem good agrees with values
$\omega_{k}^{2} a_{k}^{3}$ obtained in Kepler problem.

The substitution of the expression (\ref{2.28}) into the equality
(\ref{2.26}) yields
\begin{equation}
\label{2.30} c^{- 4}(E({\bf x}_{k}))^{2} = 1 - 2\omega_{k}^{2}
a_{k}^{2}c^{- 2}\left(1 + (1 - 4\omega_{k}^{2} a_{k}^{2} c^{-
2})^{1/2}\right)^{- 1} \approx 1 - \omega_{k}^{2} a_{k}^{2}c^{-
2},\, \, k = 1,...,9.
\end{equation}
By making use of the relations (\ref{2.16}), (\ref{2.20}), (\ref{2.21}),
(\ref{2.28}), (\ref{2.30}) we have
\begin{eqnarray}
\label{2.31} \gamma_{k} =
\left( 1 + 4\omega_{k}^{2} a_{k}^{2} c^{- 2} (1 - e_{k}^{2})^{- 1}
\left(1 + (1 - 4\omega_{k}^{2} a_{k}^{2}c^{- 2})^{1/2}\right)^{- 2}
\right)^{- 1/2} \approx \nonumber \\ 1 - 2^{- 1} \omega_{k}^{2}
a_{k}^{2} c^{- 2} (1 - e_{k}^{2})^{- 1},\, \, k = 1,...,9.
\end{eqnarray}
The value $2^{- 1}\omega_{k}^{2}a_{k}^{2} c^{- 2}(1 - e_{k}^{2})^{- 1}
\approx 1 - \gamma_{k}$ is maximal for Mercury: $1 - \gamma_{1} \approx
1.3341 \cdot 10^{- 8}$. The precession coefficients (\ref{2.31}) of the
orbits (\ref{2.15}) are practically equal to one for all planets. It agrees
with Tycho Brahe astronomical observations used by Kepler. For
a hundred years ($415$ "periods" of Mercury) the advance of Mercury's
perihelion is nearly
$(1 - \gamma_{1}) \cdot 360\cdot 415 \cdot 3600^{"} \approx 7".175$.
The relations (\ref{2.15}), (\ref{2.31}) imply the perihelion angle
\begin{equation}
\label{2.311} \phi_{k;l} \approx \phi_{k;0} + 2\pi l(1 + 2^{- 1}\omega_{k}^{2}
a_{k}^{2} c^{- 2} (1 - e_{k}^{2})^{- 1}), \, \, l = 0,\pm 1,\pm 2,....
\end{equation}
The substitution of the relations (\ref{2.20}), (\ref{2.31}) into the equality
(\ref{2.15}) yields
\begin{equation}
\label{2.33} e_{k}\cos \left( \left( 1 - 2^{- 1}\omega_{k}^{2}
a_{k}^{2} c^{- 2}(1 - e_{k}^{2})^{- 1} \right) (\phi_{k} (t) -
\phi_{k;0})\right) \approx a_{k}(1 - e_{k}^{2})r_{k}^{- 1} (t) - 1,
\end{equation}
$k = 1,...,9$.
In the general relativity the orbits (\cite{4}, Chap. 40, Sec. 40.5, relations
(40.17), (40.18)) are the approximate solutions of the geodesic equation for
the chosen metrics (\cite{4}, Chap. 40, Sec. 40.1, relation (40.3)). These
orbits are the orbits (\ref{2.33}) with the perihelion angles $\phi_{k;0} = 0$
and with the precession coefficients
$1 - 3\omega_{k}^{2} a_{k}^{2} c^{- 2}(1 - e_{k}^{2})^{- 1}$
instead of the precession coefficients
$1 - 2^{- 1}\omega_{k}^{2} a_{k}^{2} c^{- 2}(1 - e_{k}^{2})^{- 1}$. It seems that
the perihelion angles are missed in \cite{4}. We note that
$3\omega_{1}^{2} a_{1}^{2} c^{- 2}(1 - e_{1}^{2})^{- 1}\cdot 360\cdot 415 \cdot 3600^{"}
\approx 6(1 - \gamma_{1}) \cdot 360\cdot 415 \cdot 3600^{"} \approx 6\cdot 7".175 = 43".05$.
Does the orbit (\ref{2.33}), $k =1$, or the orbit (\cite{4}, Chap. 40, Sec. 40.5, relations
(40.17), (40.18)) agree with the observed Mercury's orbit? From the paper (\cite{7}, p. 361)
we know:
"Observations of Mercury are among the most difficult in positional astronomy. They have
to be made in the daytime, near noon, under unfavorable conditions of the atmosphere; and
they are subject to large systematic and accidental errors arising both from this cause
and from the shape of the visible disk of the planet. The planet's path in Newtonian
space is not an ellipse but an exceedingly complicated space-curve due to the disturbing
effects of all of the other planets. The calculation of this curve is a difficult and
laborious task, and significantly different results have been obtained by different
computers."

Substituting the relations (\ref{2.28}), (\ref{2.30}) in the
equality (\ref{2.19}) and introducing the parameter without physical
measure we get
\begin{eqnarray}
\label{2.34}  a_{k}^{- 1}r_{k}(\tau_{k}) \approx 1 + e_{k}\sin
\tau_{k}, \nonumber \\
\omega_{k} t_{k}(\tau_{k}) \approx \tau_{k} - \tau_{k;0} - e_{k}(1 -
\omega_{k}^{2} a_{k}^{2} c^{- 2})\cos \tau_{k}, \nonumber
\\ \tau_{k} = \omega_{k} a_{k} \xi_{k} \left( 1 + 2^{- 1}
\omega_{k}^{2} a_{k}^{2} c^{- 2}\right) \, \, k = 1,...,9.
\end{eqnarray}
If we neglect the values $\omega_{k}^{2}a_{k}^{2} c^{- 2}$,
then the solutions (\ref{2.33}), (\ref{2.34}) of the
equations (\ref{2.9}) coincide with the solutions of the Kepler
problem. Let us define the constant $\tau_{k;0}$ in the second
equality (\ref{2.34}) by choosing the initial time moment $t_{k}(0)
= 0$. Then the equalities (\ref{2.34}) have the form
\begin{eqnarray}
\label{2.35} a_{k}^{- 1}r_{k}(\tau_{k}) \approx 1 + e_{k}\sin \tau_{k},
\nonumber \\
\omega_{k} t_{k}(\tau_{k}) \approx \tau_{k} - e_{k}(1 -
\omega_{k}^{2} a_{k}^{2} c^{- 2})\left( \cos \tau_{k} - 1\right), \,
\, k = 1,...,9.
\end{eqnarray}
Let the direction of the first axis be orthogonal to the vector
${\bf M}({\bf x}_{1})$. Let the direction of the third axis coincide
with the direction of vector ${\bf M}({\bf x}_{3})$. Then the second
axis lies in the plane stretched on the vectors ${\bf M}({\bf
x}_{1})$ and ${\bf M}({\bf x}_{3})$. Due to the relations
(\ref{2.12})
$$
x_{1}^{1} (t) = r_{1}(t)\cos \phi_{1} (t),\, \, x_{1}^{2} (t) = -
r_{1}(t)\cos \theta_{1} \sin \phi_{1} (t),\, \, x_{1}^{3} (t) =
r_{1}(t)\sin \theta_{1} \sin \phi_{1} (t),
$$
\begin{equation}
\label{2.36} x_{3}^{1} (t) = r_{3}(t)\cos \phi_{3} (t),\, \,
x_{3}^{2} (t) = r_{3}(t)\sin \phi_{3} (t), \, \, x_{3}^{3} = 0
\end{equation}
where the inclination of Mercury orbit plane $\theta_{1} = 7^{o}$
and the values $r_{k}(t),\phi_{k} (t)$, $k = 1,3$, satisfy the
equations (\ref{2.33}), (\ref{2.35}). For the definition of Mercury
and the Earth trajectories it is necessary to define the perihelion
angles $\phi_{1;0}$, $\phi_{3;0}$ in the equations (\ref{2.33}).

"Observations of Mercury do not give the absolute position of the
planet in space but only the direction of a line from the planet to
the observer."  (\cite{7}, p. 363.) The advance of Mercury's
perihelion is given by the angle
\begin{eqnarray}
\label{2.37} \cos \alpha = \frac{({\bf x}_{1}(t_{1}(\tau_{1,1})) -
{\bf x}_{3}(t_{3}(\tau_{3,1})), {\bf x}_{1}(t_{1}(\tau_{1,2})) -
{\bf x}_{3}(t_{3}(\tau_{3,2})))}{|{\bf x}_{1}(t_{1}(\tau_{1,1})) -
{\bf x}_{3}(t_{3}(\tau_{3,1}))||{\bf x}_{1}(t_{1}(\tau_{1,2})) -
{\bf x}_{3}(t_{3}(\tau_{3,2}))|}, \nonumber \\ c(t_{3}(\tau_{3,k}) -
t_{1}(\tau_{1,k})) = |{\bf x}_{1}(t_{1}(\tau_{1,k})) - {\bf
x}_{3}(t_{3}(\tau_{3,k}))|,\, \, k = 1,2, \nonumber \\
t_{1}(\tau_{1,2}) - t_{1}(\tau_{1,1}) \leq 100T_{3} \leq
t_{1}(\tau_{1,2}) - t_{1}(\tau_{1,1}) + T_{1}
\end{eqnarray}
where the parameters $\tau_{1,1}$, $\tau_{1,2}$ are defined by
Mercury's perihelion points, the parameters $\tau_{3,1}$,
$\tau_{3,2}$ are the solutions of the second equation (\ref{2.37}),
the numbers $T_{1}$, $T_{3}$ are the orbit "periods" of
Mercury and the Earth. The quotient $T_{3}/T_{1}$ of the Earth and
Mercury orbit "periods" is approximately equal to $4.15$.

By making use of the equations (\ref{2.35}) we obtain the parameters
corresponding to Mercury's perihelion points:
\begin{eqnarray}
\label{2.39} a_{1}^{- 1}r_{1}(\tau_{1,k}) \approx  1 - e_{1},
\nonumber
\\ \omega_{1} t_{1}(\tau_{1,k}) \approx \pi \left( 2l_{k} +
3/2 \right) + e_{1}(1 - \omega_{1}^{2} a_{1}^{2} c^{- 2}),
\nonumber \\ \tau_{1,k} \approx \pi \left( 2l_{k} + 3/2
\right),\, \, k = 1,2,
\end{eqnarray}
where $l_{k}$  are the integers. The first relation (\ref{2.39})
coincides with the equality (\ref{2.22}).

According to the book (\cite{4}, Chap. 25, Sec. 25.1, Appendix 25.1)
$c^{- 1}\omega_{1} = 275.8\cdot 10^{- 17}m^{- 1}$, $c^{-
1}\omega_{3} = 66.41\cdot 10^{- 17}m^{- 1}$, $a_{1} = 0.5791\cdot
10^{11}m$, $a_{3} = 1.4960\cdot 10^{11}m$. The substitution of the second
equality (\ref{2.39}) into the third relation (\ref{2.37}) yields
$l_{2} - l_{1} = 415$.

Due to the second relation (\ref{2.37})
\begin{equation}
\label{2.40} {\bf x}_{3}(t_{3}(\tau_{3,k})) = {\bf
x}_{3}(t_{1}(\tau_{1,k})) + c^{- 1}|{\bf x}_{1}(t_{1}(\tau_{1,k})) -
{\bf x}_{3}(t_{3}(\tau_{3,k}))|{\bf v}_{3}(t_{3,k}^{\prime}),\, \, k
= 1,2.
\end{equation}
The Earth speed is small compared with the speed of light: $c^{-
1}|{\bf v}_{3}| \approx c^{- 1}\omega_{3} a_{3} \approx 0.9935\cdot
10^{- 4}$. We neglect this value ($\arcsin 10^{- 4} \approx
0^{o}.0057$). Then the relations (\ref{2.37}), (\ref{2.40}) imply
\begin{equation}
\label{2.41} \cos \alpha \approx \frac{({\bf
x}_{1}(t_{1}(\tau_{1,1})) - {\bf x}_{3}(t_{1}(\tau_{1,1})), {\bf
x}_{1}(t_{1}(\tau_{1,2})) - {\bf x}_{3}(t_{1}(\tau_{1,2})))}{|{\bf
x}_{1}(t_{1}(\tau_{1,1})) - {\bf x}_{3}(t_{1}(\tau_{1,1}))||{\bf
x}_{1}(t_{1}(\tau_{1,2})) - {\bf x}_{3}(t_{1}(\tau_{1,2}))|}
\end{equation}
where the parameters $\tau_{1,k}$, $k = 1,2$, are given by the third
relation (\ref{2.39}) and the relation $l_{2} = l_{1} + 415$.

Let us consider Mercury's perihelion points corresponding to the integers
$l_{1} = 0$ and $l_{2} = 415$. The substitution of the values corresponding
to the Mercury's perihelion, defined by the first equation (\ref{2.39}),
into the equation (\ref{2.33}) yields
\begin{eqnarray}
\label{2.42} a_{1}^{- 1}r_{1}\left( \pi \left( 2l + 3/2
\right) \right) \approx  1 - e_{1}, \nonumber \\ \phi_{1} \left(
t_{1}\left( \pi \left( 2l + 3/2 \right) \right) \right)
\approx \phi_{1;0} + 2\pi l\left( 1 + 2^{- 1}\omega_{1}^{2} a_{1}^{2}
c^{- 2}(1 - e_{1}^{2})^{- 1}\right)
\end{eqnarray}
since the value $\omega_{1}^{2}a_{1}^{2} c^{- 2} \approx 2.5509\cdot
10^{- 8}$ is negligible. We substitute the time, defined by the
second relation (\ref{2.39}), into the second relation (\ref{2.35}) for
the Earth
\begin{eqnarray}
\label{2.43} a_{3}^{- 1}r_{3}(\tau_{3} (l)) \approx 1 + e_{3}\sin
\tau_{3} (l), \nonumber \\ \omega_{3} t_{1}\left( \pi \left( 2l +
3/2 \right) \right) \approx \omega_{3} \omega_{1}^{- 1}
\left( \pi \left( 2l + 3/2 \right) + e_{1}(1 -
\omega_{1}^{2} a_{1}^{2}
c^{- 2})\right) \approx \nonumber \\
\tau_{3} (l) - e_{3}(1 - \omega_{3}^{2} a_{3}^{2} c^{- 2})\left(
\cos \tau_{3} (l) - 1\right).
\end{eqnarray}
Solving the second equation (\ref{2.43}) we get $\tau_{3} (0)
\approx 1.1748$, $\tau_{3} (415) \approx 629.09$. Substituting
these values in the first equation (\ref{2.43}) we have $a_{3}^{-
1}r_{3}(\tau_{3} (0)) \approx 1.0157$, $a_{3}^{- 1}r_{3}(\tau_{3} (415))
\approx 1.0118$. We substitute the first equation (\ref{2.43}) in the
equation (\ref{2.33}) for the Earth
\begin{equation}
\label{2.44} \cos \left( \left( 1 - \frac{\omega_{3}^{2} a_{3}^{2}
c^{- 2}}{2(1 - e_{3}^{2})}\right) \left( \phi_{3} \left( t_{1}
\left( \pi \left( 2l + 3/2 \right) \right) \right) -
\phi_{3;0} \right) \right) \approx - \frac{e_{3} + \sin \tau_{3}
(l)}{1 + e_{3}\sin \tau_{3} (l)}.
\end{equation}
The function in the right - hand side of the equation (\ref{2.44})
is monotonic with respect to the variable $e_{3}$ on the interval $0
\leq e_{3} \leq 1$. Calculating the values of this function at the
points $e_{3} = 0,1$ we get the estimation for the module of this
function which implies that the equation (\ref{2.44}) has a
solution. Substituting the solutions $\tau_{3} (l)$, $l =
0,415$, of the second equation (\ref{2.43}) in the
equation (\ref{2.44}) we get the angles in radians
$$
\phi_{3} \left( t_{1} \left( \pi \left( 2\cdot 0 + 3/2
\right) \right) \right) \approx \phi_{3;0} + 2.7521,
$$
\begin{equation}
\label{2.441} \phi_{3} \left( t_{1} \left( \pi \left( 2\cdot 415 +
3/2 \right) \right) \right) \approx \phi_{3;0} + 2.3544 +
2\pi \cdot 99 \left( 1 + 2^{- 1}\omega_{3}^{2} a_{3}^{2} c^{- 2}(1
- e_{3}^{2})^{- 1}\right)
\end{equation}
since the value $\omega_{3}^{2}a_{3}^{2} c^{- 2} \approx 0.9870\cdot
10^{- 8}$ is negligible. Substituting the radii and the angles
(\ref{2.42}), the radii (\ref{2.43}) and the angles (\ref{2.441}) in
the equations (\ref{2.36}), (\ref{2.41}) we get the equation
\begin{eqnarray}
\label{2.45}  \cos \alpha (0,415) \approx (((1 - e_{1})\cos
\phi_{1;0} - 1.0157a_{3}a_{1}^{- 1}\cos (\phi_{3;0} + 2.7521))
\nonumber
\\ \times ((1 - e_{1})\cos (\phi_{1;0} + 415\pi
\omega_{1}^{2} a_{1}^{2} c^{- 2}(1 - e_{1}^{2})^{- 1}) \nonumber
\\ - 1.0118a_{3}a_{1}^{- 1}\cos (\phi_{3;0} + 2.3544 + 99 \pi
\omega_{3}^{2} a_{3}^{2} c^{- 2}(1 - e_{3}^{2})^{- 1})) \nonumber \\
+ (0.99255(1 - e_{1})\sin \phi_{1;0} + 1.0157a_{3}a_{1}^{- 1}\sin
(\phi_{3;0} + 2.7521)) \nonumber
\\ \times ( 0.99255(1 - e_{1})\sin ( \phi_{1;0} + 415\pi \omega_{1}^{2}
a_{1}^{2} c^{- 2}(1 - e_{1}^{2})^{- 1}) \nonumber \\
+ (1.0118a_{3}a_{1}^{- 1}\sin (\phi_{3;0} + 2.3544 + 99\pi
\omega_{3}^{2} a_{3}^{2} c^{- 2}(1 - e_{3}^{2})^{- 1})) \nonumber \\
+ 0.01485(1 - e_{1})^{2}\sin \phi_{1;0} \sin (\phi_{1;0} +
415\pi \omega_{1}^{2} a_{1}^{2} c^{- 2}(1 - e_{1}^{2})^{- 1}))) \nonumber \\
\times (((1 - e_{1})\cos \phi_{1;0} - 1.0157a_{3}a_{1}^{- 1}
\cos (\phi_{3;0} + 2.7521))^{2} \nonumber \\
+ (0.99255(1 - e_{1})\sin \phi_{1;0} + 1.0157a_{3}a_{1}^{- 1}\sin
(\phi_{3;0} + 2.7521))^{2} \nonumber
\\ + 0.01485(1 - e_{1})^{2}\sin^{2} \phi_{1;0})^{- 1/2}
\nonumber \\ \times (((1 - e_{1})\cos (\phi_{1;0} + 415\pi \omega_{1}^{2}
a_{1}^{2} c^{- 2}(1 - e_{1}^{2})^{- 1}) \nonumber
\\ - 1.0118a_{3}a_{1}^{- 1}\cos ( \phi_{3;0} + 2.3544 + 99
\pi \omega_{3}^{2} a_{3}^{2} c^{- 2}(1 - e_{3}^{2})^{- 1}))^{2}
\nonumber \\ + ( 0.99255(1 - e_{1})\sin (\phi_{1;0} + 415\pi
\omega_{1}^{2} a_{1}^{2} c^{- 2}(1 - e_{1}^{2})^{- 1}) \nonumber
\\ + 1.0118a_{3}a_{1}^{- 1}\sin (\phi_{3;0} + 2.3544 + 99\pi
+ \omega_{3}^{2} a_{3}^{2} c^{- 2}(1 - e_{3}^{2})^{- 1}))^{2} \nonumber \\
+ 0.01485(1 - e_{1})^{2}\sin^{2} ( \phi_{1;0} + 415\pi \omega_{1}^{2}
a_{1}^{2} c^{- 2}(1 - e_{1}^{2})^{- 1}))^{- 1/2}.
\end{eqnarray}
The perihelion angles $\phi_{1;0}$, $\phi_{3;0}$ are needed. For the
orbits (\cite{4}, Chap. 40, Sec. 40.5, relations (40.17), (40.18)) the
perihelion angles $\phi_{k;0} = 0$. Let the perihelion angles $\phi_{1;0}$,
$\phi_{3;0}$ in the equation (\ref{2.45}) be equal to zero. Then
$\alpha (0,415) = 17^{o}.889$. According to the book
(\cite{4}, Chap. 40, Sec. 40.5, Appendix 40.3), the observed advance of
Mercury's perihelion is $1^{o}.55548 \pm 0^{o}.00011$ for a hundred years.
The angle $\alpha (0,415) = 17^{o}.889$ is not small. In our opinion for the
experimental verification of the general relativity it is necessary to obtain
the advance of Mercury's perihelion, observed from the Earth, by making use of
the Mercury and Earth orbits (\cite{4}, Chap. 40, Sec. 40.5, relations (40.17),
(40.18)) calculated without Newton gravity theory. The orbits (\cite{4}, Chap.
40, Sec. 40.5, relations (40.17), (40.18)) are the orbits (\ref{2.33}) with
the perihelion angles $\phi_{k;0} = 0$ and with the precession coefficients
$1 - 3\omega_{k}^{2} a_{k}^{2} c^{- 2}(1 - e_{k}^{2})^{- 1}$ instead of the
precession coefficients
$1 - 2^{- 1}\omega_{k}^{2} a_{k}^{2} c^{- 2}(1 - e_{k}^{2})^{- 1}$. It seems that
the perihelion angles are missed in \cite{4}. In order to calculate the advance
of Mercury's perihelion we need to know also the time dependence of the orbit
(\cite{4}, Chap. 40, Sec. 40.5, relations (40.17), (40.18)) radius.

\end{document}